\documentclass[12pt,aps,floats,showpacs]{revtex4}

\usepackage{graphicx}
\usepackage[dvips]{color}
\usepackage{colordvi}

\def\Journal#1#2#3#4{{ #1} {\bf #2}, #3 (#4) }
\def\NPA{{ Nucl. Phys.} A}

\def\PRL{Phys. Rev. Lett.}

\def\PRC{{Phys. Rev.} C}

\begin{document}
\title{Sum rules of single-particle spectral functions in hot asymmetric nuclear matter}
\author{ A. Rios and A. Polls}
\affiliation{Departament d'Estructura i Constituents de la Mat\`eria,\\
Universitat de Barcelona, E-08028 Barcelona, Spain}
\author{H. M\"uther}
\affiliation{Institut f\"ur
Theoretische Physik, \\ Universit\"at T\"ubingen, D-72076 T\"ubingen, Germany}

\begin{abstract}
The neutron and proton single-particle spectral functions in asymmetric nuclear
matter fulfill energy weighted sum rules. The validity of these sum rules
within the self-consistent Green's function approach is investigated. The various
contributions to these sum rules and their convergence as a function  of energy
provide information about correlations induced by the realistic interaction
between the nucleons. These features are studied as a
function of the asymmetry of nuclear matter.
\end{abstract}
\maketitle

\section{INTRODUCTION}
      
The microscopic description  of the single-particle properties in nuclear
matter must deal with the treatment of nucleon-nucleon (NN) correlations
\cite{muether00,di2004}. In fact, the strong short-range and tensor components
which are required in realistic NN interactions to fit the NN scattering data,
lead to corresponding correlations in the nuclear wave function. These NN
correlations are important to describe bulk properties of dense matter. They
also modify the spectral distribution of the single-particle strength in a
significant way. Recent calculations have demonstrated without ambiguity that
NN correlations produce a partial occupation of the single-particle states,
which are completely occupied in the mean field approach, and a wide
distribution  of the single-particle strength in energy. These two features
have found experimental  grounds in the analysis of $(e,e'p)$
reactions \cite{bat2001,rohe04}.

Historically, various tools have been employed to account for correlations in the nuclear
many-body wave function. These include the traditional Brueckner hole-line expansion
\cite{baldo99} and variational approaches using correlated basis functions \cite{fant98}.
Attempts have also been made to employ the technique of a self-consistent evaluation of
Green's functions (SCGF) \cite{muether00,di2004,kad62,kraeft}
to the solution of the nuclear many-body problem. This method offers various advantages:
(i) the single-particle Green’s function contains detailed information about the
spectral function, i.e., the distribution of single-particle strength as a function of
missing energy and momentum;
(ii) the method can be extended to finite temperatures, a feature which is of interest
for the study of the nuclear properties in astrophysical environments;
(iii) the Brueckner Hartree-Fock (BHF) approximation, the approximation to the hole-line
expansion which is commonly used, can be considered as a specific prescription
within this scheme.

Enormous progress in the SCGF applications to nuclear matter has been achieved in the last
years, both at zero \cite{ddnsw} and finite temperature \cite{bozek99,bozek02,frick03,frick05}.
The efforts at $T=0$ are mainly oriented to provide the appropriate theoretical support
for the interpretation of the $(e,e'p)$ experiments, while research at finite $T$ is essentially
focused on the description of the nuclear medium in astrophysical environments or on the
interpretation of heavy ion collisions dynamics.
In all cases, a key quantity is the single-particle spectral function which measures
the possibility to add or remove a particle with a given momentum at a specific energy.
A useful way to study the properties of these single-particle spectral functions is by means of
the energy weighted sum rules, which are well established in the literature and which have been
numerically analyzed for symmetric nuclear matter at zero \cite{pol94} and finite temperature
\cite{frick04}.

Recently, the single-particle spectral functions in hot asymmetric nuclear
matter have been calculated within the SCGF framework \cite{frick05}. In this
computation, the bulk properties of asymmetric dense matter were computed,
namely the energy per particle, the symmetry energy and the chemical
potentials. In addition, and since the SCGF approach gives access to the
correlated momentum distributions of neutrons and protons, the influence of the
asymmetry on the depletion of the momentum states could be studied. In
neutron-rich matter the proton states with momenta below the Fermi momentum are
more strongly depleted than in the symmetric system at the same density. In this
case the  occupation of neutron states within their Fermi sphere are enhanced
compared  to the symmetric case. This indicates that the proton-neutron
interaction is a stronger source of correlation as compared to proton-proton
and neutron-neutron interaction.

In this paper, we want to use the energy weighted sum rules to investigate the dependence
of the single-particle spectral functions on the asymmetry more in detail. 
These sum rules may help to explore the isospin dependence of the short-range 
correlations in asymmetric matter. Besides, the analysis of the energy weighted
sum rules provides valuable tests on the  numerical accuracy of our calculations.
All the computations discussed in this paper are performed in the framework of SCGF employing
a fully self-consistent ladder approximation in which the complete spectral function has been used
to describe the intermediate states in the Galistkii-Feynman equation \cite{frick03,frick05}.

After this introduction, we outline the derivation of the energy weighted sum rules
in Section 2. The results obtained for hot asymmetric nuclear using the charge dependent Bonn
potential CDBONN \cite{cdb} are presented in Section 3, where a short summary of the main
conclusions is given as well.

\section{SUM RULES FOR ASYMMETRIC NUCLEAR MATTER}

For a given Hamiltonian $H$, the single-particle Green's function for a system at finite
temperature can be defined in a grand-canonical formulation:
\begin{equation}
ig_{\tau}({\bf k}t;{\bf k'}t')= {\mathrm Tr } \left \{ \rho {\bf T }\left 
[ a_{{\bf k} \tau} (t) a_{{\bf k'}\tau}^{\dagger} (t')\right ] \right \} \, ,
\label{eq:defgr}
\end{equation}
where the subindex $\tau =n(p)$ refers either to neutrons or protons and {\bf T} is the time
ordering operator that acts on a product of Heisenberg field operators
$a_{{\bf k}\tau}(t)=e^{itH} a_{{\bf k}\tau} e^{-itH}$, in such a way that the field operator with
the largest time argument is put to the left. The trace is to be taken over all energy eigenstates
and all particle number eigenstates of the many-body system, weighted by the statistical operator,
\begin{equation}
\rho=\frac {1}{Z} e^{-\beta(H-\mu_n N_n -\mu_p N_p)} \, .
\label{eq:rho}
\end{equation}
$\beta$ denotes the inverse of the temperature and $\mu_n$ and $\mu_p$ are the neutron and proton
chemical potentials associated to a given average number of neutrons and protons. $N_n$ and $N_p$
are the operators that count the total number of neutrons and protons in the system,
\begin{equation}
N_{\tau} = \sum_{\bf k} a_{{\bf k}\tau}^{\dagger}(t) a_{{\bf k}\tau}(t) \, .
\end{equation}
The normalization factor in Eq.~(\ref{eq:rho}) is the grand partition function of statistical
mechanics,
\begin{equation}
Z= {\mathrm Tr }\,\, e^{-\beta (H-\mu_n N_n -\mu_p N_p)} \, .
\end{equation}
For a homogeneous system, the Green's function is diagonal in momentum space and depends 
only on the absolute value of ${\bf k}$ and on the difference $\xi=t'-t$. Starting from the
definition of the Green's function, we first focus on the case $\xi>0$.
In order to recover the expression in the grand canonical ensemble average of the occupation
number $n_{\tau}(k)$ for $\xi=0^+$, the following definition of the correlation functions
$g_{\tau}^{<}$ includes an additional factor $-i$ with respect to the definitions of
the Green's function $g_{\nu}$ in Eq.~(\ref{eq:defgr}),
\begin{equation}
g_{\nu}^{<}(k,\xi) = {\mathrm Tr } \left \{ \rho  e^{i\xi H} a_{{\bf k}\nu}^{\dagger}
 e^{-i \xi H}a_{{\bf k}\nu} \right \} \, .
\end{equation}
$g_{\nu}^{<}(k,\xi)$ can be expressed as a Fourier integral over the frequencies,
\begin{equation}
g_{\nu}^{<}(k,\xi) = \int_{-\infty}^{+\infty} \frac {d\omega}{2 \pi} e^{-i \omega \xi}
A_{\nu}^{<}(k,\omega)
\end{equation}
where  $A_{\nu}^{<}(k,\omega)$ is defined by
\begin{equation} 
A_{\nu}^{<}(k,\omega) = 2 \pi \sum_{rs}\frac {e^{-\beta (E_r-\mu_p N_{r(p)} -\mu_n 
N_{r(n)})}}{Z} \mid \langle \Psi_s \mid a_{k \nu} \mid \Psi_r \rangle \mid^2 
\delta( \omega -(E_r -E_s)) \, .
\label{def_spec}
\end{equation}
A similar analysis can be conducted for $\xi < 0$, yielding a function
\begin{equation}
A_{\nu}^{>}(k,\omega) =e^{\beta (\omega-\mu_\nu)} A_{\nu}^{<}(k,\omega) \, .
\end{equation}
Finally, the single-particle spectral function is defined as the sum of the two positive
functions, $A_{\nu}^{<}$ and $A_{\nu}^{>}$,
\begin{equation}
A_{\nu}(k,\omega) = A_{\nu}^{<}(k,\omega) + A_{\nu}^{>}(k,\omega) \, .
\end{equation}
In our computations, we obtain the full spectral function $A_{\nu}(k,\omega)$ and
we make use of the relations:
\begin{equation}
A^<_{\nu}(k,\omega) =  f(\omega) A_{\nu}(k,\omega) \, , \quad
A^>_{\nu}(k,\omega) =  (1-f(\omega)) A_{\nu}(k,\omega)\label{eq:specdis}
\end{equation}
to obtain $A_{\nu}^{<}$ and $A_{\nu}^{>}$, where $f(\omega)$ is the Fermi-Dirac
distribution function. The integration of the function $A_{\nu}^{<}$
leads to the occupation probability
\begin{equation}
n_{\nu}(k) = \int_{-\infty}^{+\infty} \frac{{\mathrm{d}}\omega}{2\pi}
A_{\nu}^{<}(k,\omega )\,,\label{eq:nofk}
\end{equation}
for the state with isospin $\nu$ and momentum $k$ at the inverse temperature
$\beta$.

The functions $A_{\nu}^{<}$ and $A_{\nu}^{>}$ can be compared with the hole
and particle part of the spectral function at zero temperature. In particular,
the limit of $A_{\nu}^{<}$ at zero temperature reads
\begin{equation}
A_{\nu}^{h} = 2 \pi \sum_r \mid \langle \Psi_r^{A-1} \mid a_{k\nu} \mid \Psi_0^A\rangle
\mid^2 \delta(\omega -(E_0^A -E_r^{A-1})) \, ,
\end{equation}
where $\mid \Psi_0^A\rangle$ is the ground state with $N_n$ neutrons and $N_p$ protons
(such that $A=N_p+N_n$) and $\mid \Psi_r^{A-1}\rangle $ labels the excited energy state of
a system with a neutron or a proton less (depending on which type of annihilation operator
$a_{k \nu}$ has been applied to the ground state). By its own definition, it is clear that
the lowest possible energy of the final state is the ground state energy of the $A-1$ particle
system, so that there is an upper limit for the hole spectral function
$\omega_{\nu} =E_0^A - E_0^{A-1_{\nu}} = \mu_{\nu}$,
both for neutrons and protons. In a similar way one can define the particle part
of the spectral function and find a lower bound for the excitation energy of the $A+1$
particle system, measured with respect to the ground state of the $A$ particle system. 
At zero temperature, the existence of these lower and upper bounds causes a complete
separation in energy between the particle and the hole part of the spectral function.

In the $T$ matrix approximation to the self-energy reported in Ref.\cite{frick05}, one
can determine the single-particle Green's function as the solution of Dyson's equation for
any complex value of the frequency variable $z$
\begin{equation}
g_{\nu}(k,z) = \frac {1}{z - \frac {k^2}{2m} - \Sigma_{\nu}(k,z)} \, .
\label{eq:green1}
\end{equation}
The analytical properties of the finite temperature Green's function allows one to derive
the corresponding Lehmann representation, which for slightly complex values of the
frequency can be written as
\begin{equation}
g_{\nu}(k,\omega+ i \eta) =  \int_{-\infty}^{+\infty} \frac {d\omega'}{2 \pi} 
 \frac {A(k,\omega')}
{\omega - \omega' + i \eta} \, .
\label{eq:green2}
\end{equation}

The sum rules for the spectral functions can be derived from the asymptotic behavior
at large $\omega$ by expanding the real part of both previous expressions for the Green's
function, Eqs.
(\ref{eq:green1}) and (\ref{eq:green2}), in powers of $1/\omega$. This yields
\begin{equation}
\mbox{Re } g_{\nu}(k,\omega) = \frac {1}{\omega} \left \{ 1 + \frac {1}{\omega} \left [ \frac {k^2}
{2m} + \lim_{\omega \rightarrow \infty} {\rm Re} \Sigma_{\nu}(k,\omega \right ] + ... \right \}
\end{equation}
and 
\begin{equation}
\mbox{Re } g_{\nu}(k,\omega) = \frac {1}{\omega} \left \{
 \int_{-\infty}^{+\infty} d\omega' A_{\nu}(k,\omega') 
 +\frac{1}{\omega} \int_{-\infty}^{+\infty} d\omega' \omega' A_{\nu}(k,\omega') + ... \right \} \, .
\end{equation}

By comparing the first two expansion coefficients, one finds the $m_0$
\begin{equation}
\int_{-\infty}^{+\infty} \frac{{\mathrm{d}}\omega}{2\pi}
A_{\nu}(k,\omega)=1,
\end{equation}
and the $m_1$ sum rules
\begin{equation}
\int_{-\infty}^{+\infty} \frac{{\mathrm{d}}\omega}{2\pi} \omega
A_{\nu}(k,\omega)=\frac{k^2}{2m}+\lim_{\omega\rightarrow\infty}
{\mathrm{Re}} \Sigma_{\nu}(k,\omega) \, .
\label{eq_m1}
\end{equation}

Is is worth to mention that in our scheme the self-energy is derived in the $T$~matrix
approximation and so its real part is computed from the imaginary part using the
following dispersion relation:
\begin{equation}
\label{eq_real_sigma}
{\mathrm{Re}}\,\Sigma_{\nu}(k,\omega)=\Sigma_{\nu}^{\infty}(k)-\frac{{\mathcal{P}}}{\pi}
\int_{-\infty}^{+\infty}
{\mathrm{d}}\lambda\,
\frac{{\mathrm{Im}}\Sigma_{\nu}(k,\lambda+{\mathrm{i}}\eta)}{\omega-\lambda} \, .
\end{equation}
In the derivation of the previous equation, the spectral decomposition of the Green's
function is already used, so it is a property of the $T$~matrix approach that it fullfils
the sum rule. Nevertheless, the sum rules still provide a useful consistency check for
the numerics. The first term on the right hand side of Eq.~(\ref{eq_real_sigma}) is the
energy independent part of the self-energy,
\begin{equation}
\label{eq_HF}
\Sigma_{\nu}^{\infty}(k)=
\sum_{\tau}
\int \frac{{\mathrm{d}}^3k^{\prime}}{(2\pi)^3}
\left<{\mathbf{k}}\nu {\mathbf{k}}^{\prime}\tau \right|V
\left|{\mathbf{k}}\nu {\mathbf{k}}^{\prime}\tau \right>_A n_{\tau}({\mathbf{k}}^{\prime})
\end{equation}
which can be identified with the limit
 $\lim_{\omega \rightarrow \infty}\mathrm{Re} \Sigma_{\nu}(k,\omega)$,
since the dispersive part decays like $1/\omega$ for $\omega \rightarrow \pm \infty$ .
Eq.~(\ref{eq_HF}) looks like a Hartree-Fock (HF) potential. However, $n_{\nu}(k)$
is the momentum distribution of Eq.~(\ref{eq:nofk}) containing the depletion effects due to NN correlations
and temperature.

\section{RESULTS AND DISCUSSION}

All the results discussed in this paper have been obtained with the charge-dependent
Bonn (CDBONN) potential.
Since we want to focus our study on the dependence of the observables on asymmetry,
we will consider only one density
$\rho=0.16$ fm$^{-3}$ and one single temperature $T=5$ MeV. This temperature is 
low enough to allow for conclusions in the limit of
zero temperature but high enough to avoid instabilities associated with
neutron-proton pairing \cite{von93,alm96,bo99}. As for the proton fractions, we will 
consider three different cases. The first one corresponds to symmetric nuclear matter
$x_p=\rho_p/\rho=0.50$ and serves as a guide-line for the other cases. The last one has
a very low proton fraction $x_p=0.04$ and corresponds to the $\beta-$stable composition
of matter with nucleons, electrons and muons at $T=5$ MeV and $\rho=0.16$ fm$^{-3}$. This
composition can be computed thanks to the fact that we know the asymmetry dependence of both the
neutron and the proton chemical potentials at this density \cite{frick05}. Finally,
we consider an intermediate fraction $x_p=0.30$ which is useful to identify the effects
for low asymmetries of the system.

\begin{figure}
\begin{center}
\includegraphics[height=12.5cm]{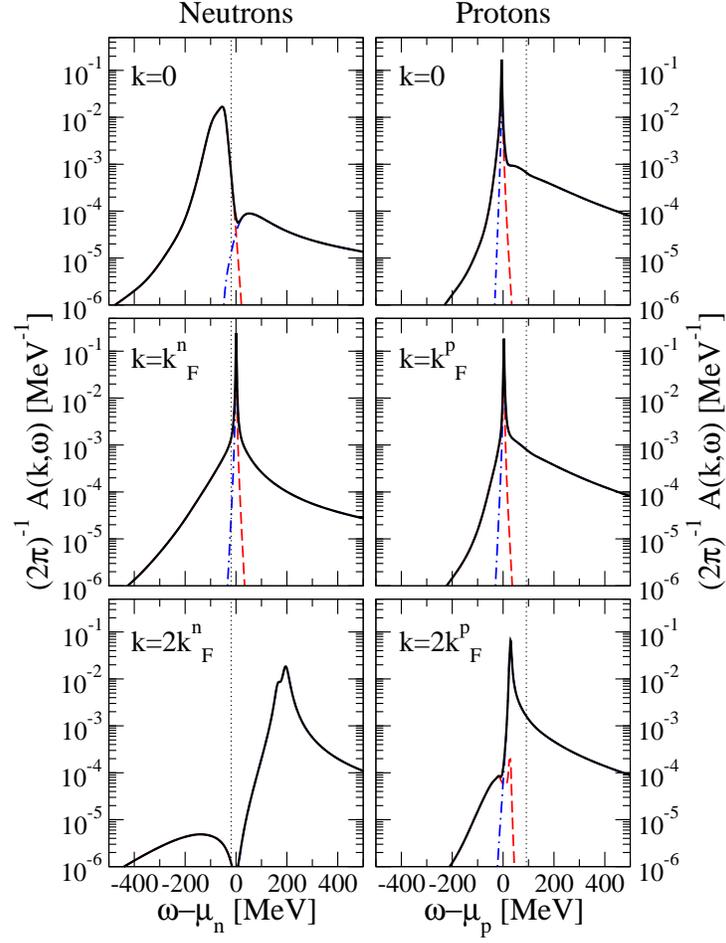}
\end{center}
\caption{(Color online) Neutron (left panels) and proton (right panels) single-particle spectral
functions at $\rho=0.16$ fm$^{-3}$, $T=5$ MeV and proton fraction $x_p=0.04$
for three different momenta. Both $A_{\nu}^{<}$ (dashed lines)
and $A_{\nu}^{>}$ (dot-dashed lines) are displayed together with the
total spectral function $A_{\nu}$ (solid lines).}
\label{fig:01}
\end{figure}

We start by discussing the momentum dependence of the single-particle spectral functions
of neutrons and protons. 
Depending on how far above or below the considered momentum is from the Fermi momentum of
the corresponding particle, we expect a very different behavior for the spectral function.
Fig.~\ref{fig:01} shows the neutron (left panels) and proton (right panels) single-particle
spectral functions for three different momenta ($k=0,k_F^{\nu}$ and $2 k_F^{\nu}$, with
$k_F^{\nu}$ the Fermi momentum of each nucleon species) at a proton fraction of
$x_p=0.04$.
The dotted vertical line corresponds to $\tilde \omega \equiv \omega-\mu_{\nu} = -\mu_{\nu}$ and
indicates the point below which
the contribution to the sum rule $m_1$ in Eq.~(\ref{eq_m1}) is negative. 

The contributions $A^<_{\nu}$
and $A^>_{\nu}$ are given by the dashed and dash-dotted lines, respectively.
Since we are considering finite temperature, thermally excited states are always included
in the definition of the spectral function Eq.~(\ref{def_spec}). This leads to
contributions to $A^<_{\nu}$ at energies $\omega$ larger than $\mu_{\nu}$
($\tilde \omega$ larger than zero). Similarly,  $A^>_{\nu}(k,\omega)$ extends to the region 
below $\mu_{\nu}$.
In general there is no longer a clear separation in energy between
$A_{\nu}^>$ and $A_{\nu}^<$, as it is the case for $T=0$. Actually, the maxima of
$A_{\nu}^>$ and $A_{\nu}^<$ can even coincide.

For the case of neutrons at $k=0$, the peak of the spectral function is provided
by $A_n^{<}$. This is due to the fact that the position of this peak, which can
be identified with the quasi-particle energy $\epsilon_{qp}^n(k)$, is well below
the chemical potential $\mu_n$, which implies that the value of Fermi-Dirac distribution
function $f(\omega)$ at this energy $\omega=\epsilon_{qp}^n(0)$ is very close to
one (see Eq.~\ref{eq:specdis}). Since the quasi-particle peak is well below
$\mu_n$, the thermal effects do not fill up the minimum in the spectral function
at $\omega=\mu_n$ and we still observe some kind of separation in energy between
$A_n^{<}$ and  $A_n^{>}$. However, around $\tilde \omega =0$ there is an
energy interval, in which both $A_n^{<}$ and $A_n^{>}$ are small but different
from zero, i.e. an interval in which these functions overlap.

A similar situation is observed for the neutron spectral function at $k=2
k_F^n$. In this case the quasi-particle energy is well above $\mu_n$, which
means that $(1-f(\omega))$ is very close to one at this energy and the peak 
structure is supplied by $A_n^{>}$ (see Eq.~\ref{eq:specdis}). Therefore we
observe a clear separation between the hole part ($A_n^{>}$) and the particle
part ($A_n^{<}$) of the spectral function even at finite temperature.

At $k=k_F^n$, the quasi-particle peak is close to the chemical potential.
This means that $f(\epsilon_{qp}^n(k))$ is around 0.5 and therefore $A_n^{<}$ 
and $A_n^{>}$ have a peak at $\tilde \omega=0$. The thermal effects fill up the
zero that the spectral function has at $\tilde \omega=0$ in the $T=0$ limit and the
overlap region of  $A_n^{<}$ and $A_n^{>}$ is enhanced.

In the case of very small proton fraction, like $x_p$ = 0.04, the Fermi
momentum $k_F^p$ is rather small and consequently the quasi-particle energies
are close to the Fermi energy $\mu_p$ for $k=k_F^p$ as well as $k=0$ and $k=2
k_F^p$. Therefore, although the relative distances to the Fermi surfaces are the
same, there are strong overlaps between $A_p^<$ and $A_p^>$ at all the three
momenta. Even in the case of $k=2 k_F^p$, a small peak structure for $A_p^<$ around
$\tilde \omega=0$ can be observed.

In Table \ref{tab:01} (neutrons) and Table \ref{tab:02} (protons) we report the fraction of
the integrated strength of $A_{\nu}^{<}$ below and above the corresponding chemical
potential, $\mu_{\nu}$. In fact, if we identify $A_{\nu}^{<}$ with the $T=0$ hole
spectral function, the integrated strength above the chemical potential would be exactly
zero (since $A_{\nu}^{<}=0$ for $\omega > \mu_{\nu}$). In this sense, the integrated
strength above the chemical potential can be considered as a genuine thermal effect.
As expected, these thermal effects
are important around $k_F^{\nu}$, where the overlap between $A_{\nu}^{<}$ and
$A_{\nu}^{>}$ is significant. Notice also that for this very neutron-rich system,
neutrons are less affected by temperature, while protons (which have a substantially
lower Fermi momentum and can thus be considered as a dilute system) are much more
influenced by temperature. For instance, protons show a large amount of strength above
$\mu_F^{p}$ up to momenta $k/k_F^p \sim 2$. For the sake of
completeness, the corresponding occupation numbers for each species are also
listed in those tables.
 
\begin{figure}
\begin{center}
\includegraphics[height=8.5cm]{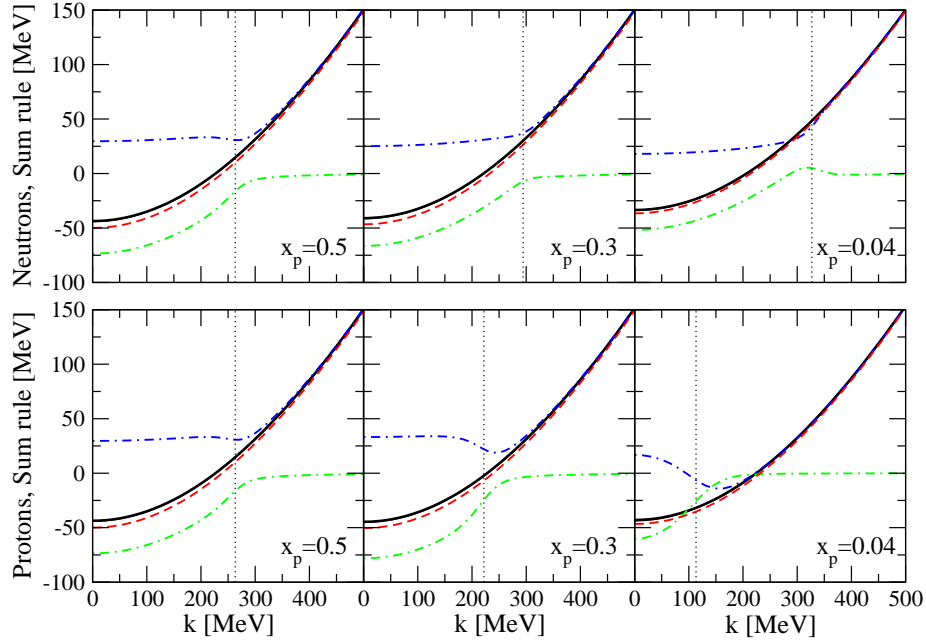}
\end{center}
\caption{(Color online) Energy weighted sum rule $m_1$ (solid lines) for neutrons (panels above)
and protons (panels below) at $\rho=0.16$ fm$^{-3}$, $T=5$ MeV and several
proton fractions. Both the right- and the left-hand sides of
Eq.~(\protect{\ref{eq_m1}}) are displayed, but
the sum rule is so well fulfilled that they cannot be distinguished.
The contributions to $m_1$ that come from $A_{\nu}^{<}$ and $A_{\nu}^{>}$
are indicated by the upper and the lower dash-dotted lines, respectively. The
result for the energy weighted sum rule obtained from the Hartree-Fock
approximation is represented by the dashed lines.}
\label{fig:02}
\end{figure}

The $m_0$ sum rule is fulfilled with an accuracy better than $.1$ \% for both 
neutrons and protons in the
complete momentum range. Results for $m_1$ both for neutrons (upper panels) and
protons (lower panels) are reported in Fig.~\ref{fig:02} at the three proton fractions
$x_p=0.5,0.3$ and $0.04$. The vertical dotted lines indicate the location of
$k_F^{\nu}$ for each case. Since the sum rule is satisfied better than $1$ \% in all
cases, the left- and the right-hand sides of Eq.~(\ref{eq_m1}) (solid lines) lie on
top of each other and cannot be distinguished. To understand the $k$-dependence of
the different contributions to $m_1$, it is useful to keep in mind the proton and neutron
chemical potentials at the different concentrations given in Table \ref{tab:03}.

The lower dash-dotted line shows the $m_1$ contribution from $A_{\nu}^{<}$.
For momenta below $k^{\nu}_F$, the contribution of $A_{\nu}^{<}$ to $m_1$ is dominated by
the quasi-particle peak, which lies below $\mu_{\nu}$. As the momentum increases, the
peak appears closer to $\omega=0$ and its weight in the integral is diminished,
which makes the integral smaller in magnitude and thus the contribution to
the sum rule becomes an increasing function of $k$. When we get closer to the Fermi
momentum $k^{\nu}_F$, the peak moves to $\omega \sim \mu_{\nu}$ and the
position of the chemical potential becomes crucial. As far as the chemical
potential is negative, the contribution of $A_{\nu}^{<}$ will be negative because the
Fermi-Dirac function falls off close to $\mu_{\nu}$ and the integral will only catch
a little positive zone if the chemical potential is close to $\omega \sim 0$. If, on
the other hand, the chemical potential is positive, the integrand is not zero for 
$\omega > 0$ and there can be a non-negligible positive contribution to the integral when
the quasi-particle peak lies between $\omega = 0$ and $\omega = \mu_{\nu}$.
This cancellations between positive and negative contributions give
rise to the structures observed for the $A_{\nu}^{<}$ sum rule close to the Fermi
momentum. In the upper-right panel of Fig.~\ref{fig:02}, for instance, the asymmetry
is so extreme that the neutron chemical potential $\mu_n>0$ and the positive contribution
to the integral is important enough to pull the sum rule to positive values
for $k \sim k^n_F$. Finally, for high momenta, $A_{\nu}^{<}$ is strongly suppressed
(the quasi-particle peak lies in the region $\omega > \mu_{\nu}$, which is suppressed by
the Fermi-Dirac factor) and the contribution to the sum rule goes to zero.

The upper dash-dotted line displays the contribution from $A_{\nu}^{>}$. Due to the
short-range correlations, there is always a high energy tail that gives rise
to a positive contribution which, for momenta well below $k_F^\nu$,
is nearly constant. In the case of the
neutron sum rules, this constant decreases slightly with decreasing proton fraction,
although the chemical potential $\mu_n$ increases significantly.
Such a small decrease is in accordance with the fact that the neutron-proton
correlations are dominant compared to the neutron-neutron ones.
These high energy tails (which are mainly caused by short-range correlations)
are, nevertheless, not much affected by asymmetry.
When the momenta becomes larger than $k_F^n$, $m_1$ increases monotonously following
the location of the quasi-particle peak, which moves to higher energies when the
momentum grows.

Also for protons one observes a contribution from $A_{\nu}^{>}$ to the
sum rule which is almost a constant for momenta smaller than the Fermi
momentum. In this case, however, this constant is almost independent of the
proton fraction, although the chemical potential $\mu_p$ gets significantly more
attractive with decreasing $x_p$.
At small values of $x_p$, the neutron abundance is large and this leads to strong
correlation effects in the proton spectral function.
In particular at $x_p=0.04$, which is the case corresponding to the lowest Fermi momenta,
the total contribution of $A_p^{>}$ is the result of a balance between positive
and negative contributions. These can result in a minimum for $k \sim k_F^p$.
In that particular proton fraction (see Fig.~\ref{fig:01}), the quasi-particle
peaks lies always below $\omega=0$ and thus it gives rise to a negative contribution. The
relative width of this peak and the detailed structure of the high-energy tail can push the
integral to negative values (almost $-15$ MeV at $k \sim 160$ MeV). At larger
momenta, however, the contribution to $m_1$ of $A_p^>$ starts to grow steadily. Notice
that for this asymmetry we can not say that this is due to the movement of the quasi-particle
peak, because even at $k=2k^p_F$ the quasi-particle peak lies in the region where the contribution to the
integral is negative. The growth in $k$ is, to a large extent, a consequence of
the contributions to the spectral functions at large energies.

In addition to these contributions, we have plotted in the dotted lines of Fig.~\ref{fig:02}
the Hartree-Fock approximation of $m_1$ at the same temperature, density and proton
fraction. This approximation turns out to give a very good estimate of $m_1$, an interesting
result which has been already observed in the previous analysis for symmetric
nuclear matter \cite {pol94,frick04}. This result allows for a quantitative estimate
of the amount of correlations produced by a given NN potential without performing
sophisticated many-body calculations.

Furthermore, it is worth noticing that the value of $m_1$ (specially for large $k$) is not so much
affected by the asymmetry of the system. It is very similar for both neutrons and protons
in all the three cases considered. This is a somehow surprising result, since we have
seen that the separate contributions of $A_{\nu}^<$ and $A_{\nu}^>$ can indeed be very different.
Of course, at high momentum the kinetic energy is dominant and this
could explain this fact in part. 

\begin{figure}
\begin{center}
\includegraphics[height=12cm,angle=-90]{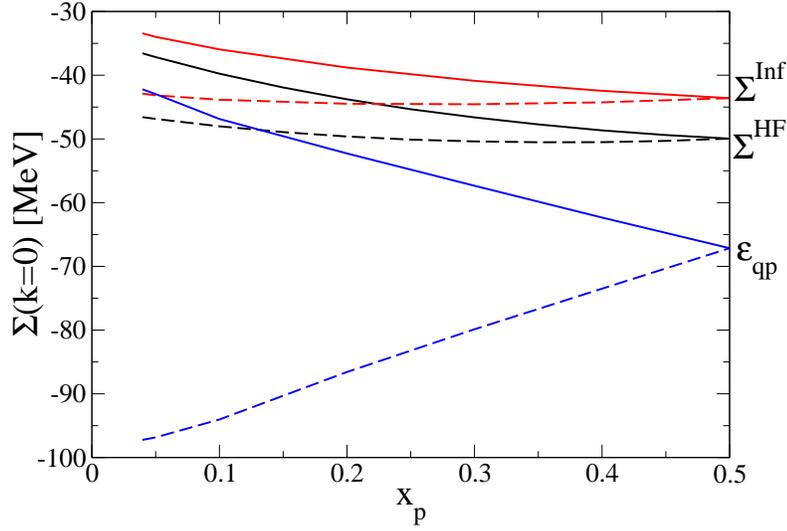}
\end{center}
\caption{(Color online) $\Sigma_{\nu}^{\infty}$, $\Sigma_{\nu}^{HF}$ and $\epsilon_{qp}^{\nu}$
at zero momentum as a function of the proton fraction at $\rho=0.16$ fm$^{-3}$
and $T=5$ MeV for neutrons (solid lines) and protons (dashed lines).}
\label{fig:03}
\end{figure}

However, even at zero momentum the effect of the asymmetry
is rather moderate, as we can see from Fig.~\ref{fig:03}, where $\Sigma_{\nu}^{\infty}$,
$\Sigma_{\nu}^{HF}$ (the Hartree-Fock approximation to the self-energy)
and the quasi-particle energy $\epsilon_{qp}^{\nu}$
at zero momentum are plotted as a function of the proton fraction.
Let us consider the highly asymmetric case of $x_p=0.04$.
The isospin splitting of the Hartree-Fock energies ($k=0$) at this asymmetry is
around 10 MeV, which is rather small compared to the isospin splitting of the
quasi-particle energies, which is around 55 MeV. This means that the stronger
binding of the proton states as compared to the neutron states is to a small extent
due to the attractive neutron-proton interaction in the bare NN potential, the
Born term of the $T$-matrix. The obtained attraction is then mainly due to the terms
in the $T$-matrix coming from the second and higher orders in the NN potential.
In other words, it is an effect of strong correlations in the isospin equal to zero
NN channels (like the $^3S_1-^3D_1$ partial wave), which lead to the deeply bound quasi-particle
energy for protons at this large neutron abundance. These correlation effects,
however, are also important for a redistribution of single-particle strength for
protons with $k=0$ to energies well above the Fermi energy, leading to a low
occupation probability (see Table \ref{tab:02}) and a large positive
contribution to the sum rule $m_1$. This positive contribution yields to a value
of $\Sigma^{\infty}$, which is even above the HF energy. The effects of
correlations for the corresponding energies of the neutron state are much
weaker, indicating that correlation effects in the isospin one NN channels are
weaker.

So we observe a compensation of correlation effects in the energy weighted
sum rule. On the one hand, correlations lead to a more attractive quasi-particle energy (compared
to the HF result) and therefore to a more attractive contribution of the 
quasi-particle pole to the sum rule. On the other hand, correlations shift
single-particle strength to high energies, which yield a repulsive contribution
to $m_1$. This compensation of correlation effects explains that the value of
the energy integrated sum rule for the correlated system can be very similar to
the HF result.

\begin{figure}
\begin{center}
\includegraphics[height=12cm,angle=-90]{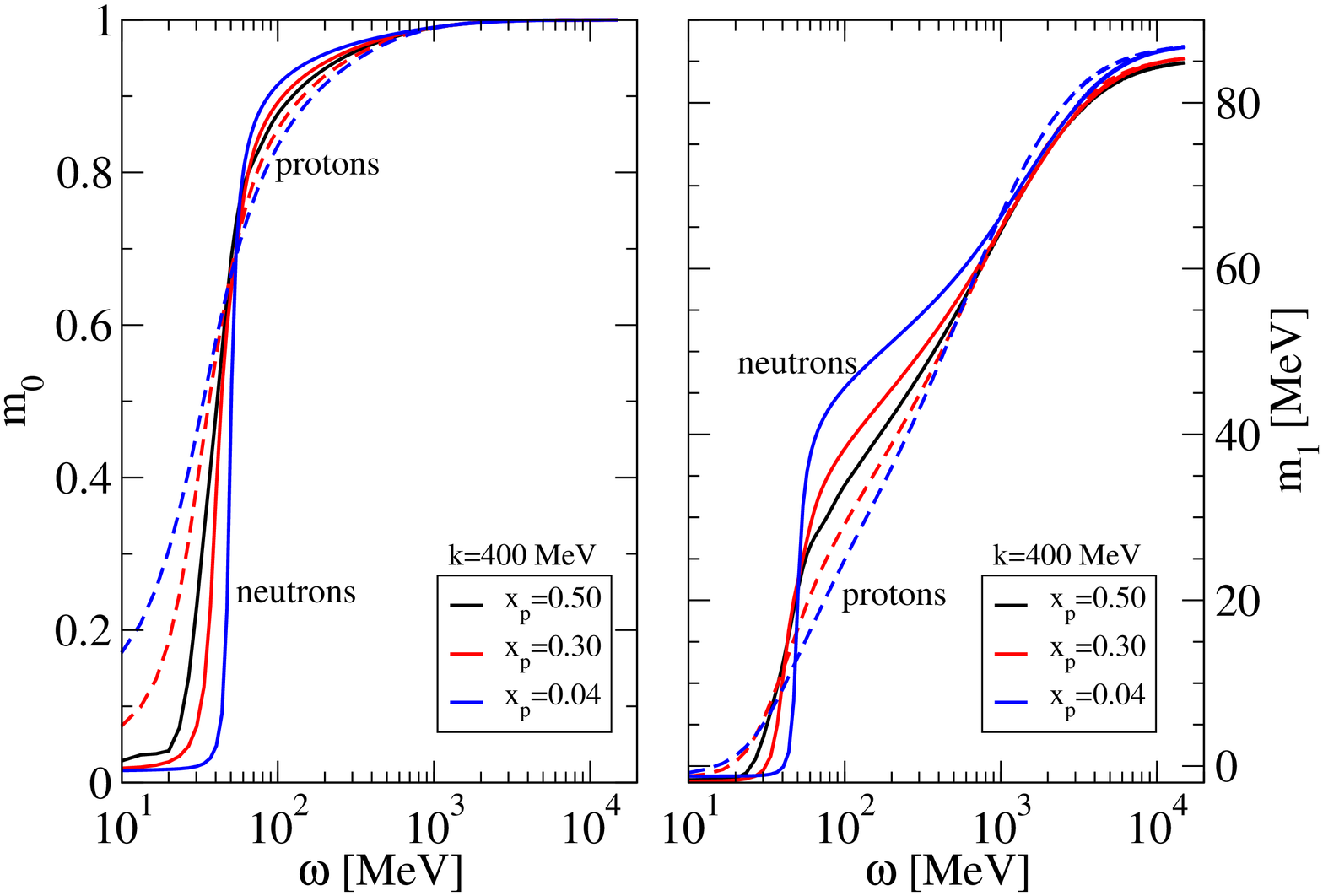}
\end{center}
\caption{(Color online) Saturation of the sum rules $m_0$ (right panel) and $m_1$ (left
panel) for neutrons (solid lines) and protons (dashed lines) at three
different proton fractions.
The momentum is k=400 MeV and the density and temperature, the same
 same as in the previous figures.}
\label{fig:04}
\end{figure}

This compensation of correlation effects can also be observed in
Fig.~\ref{fig:04}, which shows the exhaustion of the sum rules $m_0$ (left panel) and $m_1$
(right panel) at a momentum $k=400$ MeV for both neutrons (solid lines)
and protons (dashed lines), at the three different proton fractions. The first thing to be
observed is that the asymmetry affects mainly the amount of strength exhausted at 
intermediate energy regions, while there is not a noticeable difference at high energies.
For $x_p=0.5$, i.e. the symmetric case, the quasi-particle energies for neutrons and protons
at $k=400$ MeV, are the same: $\epsilon_{qp}=36.45$ MeV. When the proton fraction decreases,
the quasi-particle energies split. Neutrons become more repulsive
($\epsilon_{qp}^{n}=41.88$ MeV), in contrast with protons ($\epsilon_{qp}^{p}=30.86$ MeV).
This behavior is confirmed at $x_p= 0.04$, resulting in $\epsilon_{qp}^{n}=
49.35$ MeV and  $\epsilon_{qp}^{p}=25.22$ MeV.
In the case of $m_0$, since the quasi-particle peak of protons is located at lower energies,
the amount of strength exhausted at lower energies is larger for protons than for neutrons.
This trend changes at intermediate energies and both strengths merge together for large energies,
which is an indication of the fact that, for these momenta and asymmetries, the quasi-particle
contributions to the sum rule of neutrons are larger than those of protons, thus explaining
why the increase in $m_0$ at the quasi-particle pole is larger in the former case.
The same type of analysis is valid for $m_1$. Notice that, as previously mentioned, the final
value of $m_1$ is not much affected by the asymmetry and the behavior at high energies
is the same for all the cases.

To summarize, we have analyzed the behavior of the energy weighted sum rules of 
single-particle spectral functions of hot asymmetric nuclear matter. The sum rules are
very well fulfilled, because the $T$-matrix approximation itself respects the analytical properties
of both the self-energy and the Green's function, in which the sum rules are based. 
Nevertheless, they are a good test of the numerical consistency of the calculation, which may be helpful 
e.g. in deciding the best distribution
of the energy mesh points when one has to work with spectral functions. 

Employing realistic NN interactions, an important source of correlations come from
the strong components in the neutron-proton interaction. As a consequence,
one observes in neutron-rich matter a larger depletion of the occupation
probabilities for protons with momenta below the Fermi-momentum than for
neutrons. One also finds that these correlations, at this asymmetry, lead to much
more attractive quasi-particle energies for protons than those obtained in the HF
approximation. The same effect is observed in neutrons, although it is considerably weaker.
This shift of the quasi-particle energies tends to lead to more attractive energy weighted
sum rules $m_1$. This effect is compensated by the fact that the very same
correlations are also responsible for a shift of single-particle strength to
high positive energies. As a consequence, the energy integrated sum rules
and the isospin splitting in asymmetric nuclear matter for the
correlated system yield results which are very close to the HF ones.
In both cases, however, there is not a strong dependence on asymmetry
due to the lack of tensor correlations.
This relocation of single-particle strength, on the other hand, can be nicely observed in the
convergence of the energy weighted sum rule, which is therefore an indicator of
correlation effects.

The authors are grateful to Prof. Angels Ramos and to Dr. Tobias Frick
for enlightening and fruitful discussions. 
We would like to acknowledge financial support from the {\it Europ\"aische
Graduiertenkolleg T\"ubingen - Basel} (DFG - SNF).
This work is partly supported by DGICYT contract BFM2002-01868 and 
the Generalitat de Catalunya contract SGR2001-64.
One of the authors (A. R.) acknowledges the support of DURSI and the European Social Funds.

\newpage

\begin{table}[t]
\begin{displaymath}
\begin{array}{c|ccc|}
k/k_F^n  &  \mbox{below } \mu_n \, [\%] &  \mbox{above } \mu_n \, [\%] & n(k)  \\ \hline \hline
 0.0     &   99.98                      &   0.02                        & 0.971   \\
 0.5     &   99.96                      &   0.04                        & 0.960  \\
 1.0     &   55.58                      &   44.42                       & 0.464   \\
 1.5     &   98.48                      &   1.52                        & 0.006 \\
 2.0     &   99.76                      &   0.24                        & 0.001 \\ \hline

\end{array}
\end{displaymath}
\caption{Strength distribution of $A^<_n$ at $\rho=0.16$ fm$^{-3}$, $T=5$ MeV and proton
fraction $x_p=0.04$.  The numbers give the fraction of the integrated
strength above and below the neutron chemical potential $\mu_n$.
The last column reports the occupation of the respective neutron momentum state.}
\label{tab:01}
\end{table}

\begin{table}[t]
\begin{displaymath}
\begin{array}{c|ccc|}
k/k_F^p  &  \mbox{below } \mu_p \, [\%] &  \mbox{above } \mu_p \, [\%] & n(k)  \\ \hline \hline
 0.0     &   97.62                      &   2.38                        & 0.605   \\
 0.5     &   94.88                      &   5.12                        & 0.531  \\
 1.0     &   21.62                      &   78.38                       & 0.280   \\
 1.5     &   24.26                      &   75.74                       & 0.056 \\
 2.0     &   64.10                      &   35.90                        & 0.009 \\ \hline

\end{array}
\end{displaymath}
\caption{Strength distribution of $A^<_p$ at $\rho=0.16$ fm$^{-3}$ $T=5$ MeV and proton 
fraction $x_p=0.04$. The numbers give the fraction of the integrated
strength above and below the proton chemical potential $\mu_p$.
The last column reports the occupation of the respective proton momentum state.}
\label{tab:02}
\end{table}

\begin{table}[t!]
\begin{displaymath}
\begin{array}{c|cc}
 x_p      &  \mu_n \mbox{ [MeV]}         &  \mu_p \mbox{ [MeV]}  \\ \hline
 0.50     &   -23.46                     &   -23.46   \\
 0.30     &   -2.78                      &   -48.33   \\
 0.04     &   19.21                      &   -91.45   \\
\end{array}
\end{displaymath}
\caption{Neutron and proton chemical potentials at $\rho=0.16$ fm$^{-3}$,
 $T=5$ MeV and different proton fractions.}
\label{tab:03}
\end{table}

\newpage

\begin{thebibliography}{99}
\bibitem{muether00} H. M\"uther and A. Polls, \Journal{Prog. Part. Nucl.
Phys.}{45}{243}{2000}.

\bibitem{di2004} W.H. Dickhoff and C. Barbieri, \Journal{Prog. Part. Nucl.
Phys.}{52}{377}{2004}.

\bibitem{bat2001} M.F. Batenburg, Ph.D thesis, University of Utrecht, 2001.

\bibitem{rohe04} D. Rohe, \textit{et al.}, \Journal{\PRL}{93}{182501}{2004}.

\bibitem{baldo99} M. Baldo, {\em
Nuclear methods and the Nuclear Equation of State}, Int. Rev. of Nucl.
Physics, Vol. 9 (World-Scientific, Singapore, 1999).

\bibitem{fant98} S. Fantoni and A. Fabrocini in {\em Microscopic Quantum
Many-Body Theories and Their Applications}, eds. J. Navarro and A. Polls
(Springer 1998).

\bibitem{kad62} L. P. Kadanoff and G. Baym,
{\em Quantum Statistical Mechanics} (Benjamin, New York, 1962).

\bibitem{kraeft} W. D. Kraeft, D. Kremp, W. Ebeling and G. R\"opke, {\em
Quantum Statistics of Charged Particle Systems} (Akademie-Verlag,
Berlin, 1986).

\bibitem{ddnsw} Y. Dewulf, W.H. Dickhoff, D. Van Neck, E.R. Stoddard,
and M. Waroquier, \Journal{\PRL}{90}{152501}{2003}.

\bibitem{bozek99} P. Bo\.zek, \Journal{\PRC}{59}{2619}{1999}.

\bibitem{bozek02} P. Bo\.zek, \Journal{\PRC}{65}{054306}{2002}.

\bibitem{frick03} T. Frick and H. M\"uther, \Journal{\PRC}{68}{034310}{2003}.

\bibitem{frick05} T. Frick, H. M\"uther, A. Rios, A. Polls and A. Ramos, \Journal {\PRC}
{71}{014313}{2005}.

\bibitem{pol94} A. Polls, A. Ramos, J. Ventura, S. Amari and
W.H. Dickhoff, \Journal{\PRC}{49}{3050}{1994}.

\bibitem{frick04} T. Frick, H. M\"uther, and  A. Polls, \Journal {\PRC}
{69}{054305}{2004}.

\bibitem{cdb} R. Machleidt, F. Sammarruca, and Y. Song, \Journal {\PRC}{53}{R1483}{1996}.

\bibitem{von93} B. E. Vonderfecht, W. H. Dickhoff, A. Polls, and A. Ramos, 
\Journal {\NPA}{555}{1}{1993}.

\bibitem{alm96} T. Alm, G. R\"opke, A. Schnell, N. H. Kwong, and H. S. K\"ohler,
\Journal {\PRC}{53}{2181}{1996}.

\bibitem{bo99} P. Bo\.zek, \Journal {\NPA}{657}{187}{1999}.
\end{thebibliography}
\end{document}